# A first order phase transition and self-organizing states in unidomain ferromagnet

**V. V. Nietz**

*Joint Institute for Nuclear Research,
Dubna, Moscow Region 141980, Russia*
E-mail address: nietz@jinr.ru

**Abstract**

Unidomain ferromagnet during the first order phase transition induced by a magnetic field directed along the axis of anisotropy, has been analyzed. It is shown that there is continuous spectrum of stationary states, which depends of relation between acting field value and the demagnetisation field.

At addition action of the periodic field perpendicular to the main magnetic field with a certain frequency, there are dynamic equilibrium state, ie "self-organizing states" of ferromagnet, when the entropy increase connected with dissipation is compensated by the negative entropy flow due to the periodic field.

It is shown that under these conditions, by varying the frequency of the periodic field, we can control the self-organising system, ie decrease or increase the system energy and, correspondingly, change the direction of magnetisation in ferromagnet.

## 1. Introduction

A very large number of phenomena and processes are known, which can be classified as "self-organizing systems" or "dissipative structures" [1-4]. Processes belonging to this category are, for example, sounding of wind and stringed musical instruments, a whistle sound, existence of proteins, development of plants, functioning of animals and humans. Generally, the life itself in all its forms is an example of such "self-organizing systems". It may seem surprising that, unlike nature, the man himself was able to invent so limited number of such systems. This could include such examples that can be reproduced on the laboratory table: the chemical "Zhabotinskiy reaction" [5, 6], "Benar cells" at liquid boiling [7]. Precessing ball solitons during the magnetic phase transition in ferromagnet could also be considered as "a self-organizing system" or "self-organizing states" (SOS) [8]. Some of these systems are structures periodic in space or in time. Others are more complex. But the common feature of all these processes is that the loss of energy in the system associated with dissipation, is fully offset by the influx of energy from external sources, i.e. inflow of entropy due to the dissipation is compensated by the negative flow of entropy due to the coupling to an external source.

In the second part of this article, the first-order phase transition in a unidomain ferromagnet has been analyzed. The sole purpose of the unidomain condition is to exclude extraneous sources of nucleation of a new phase, such as domain walls or external boundaries of the crystal. (For example, $2\pi$-degree boundaries themselves are nuclei of a new phase.) Under such conditions, the process of ferromagnetic changes is considered to depend on the magnitude of applied field and the influence of demagnetizing field.

Analysis scheme of ferromagnetic is presented in Figure 1. Initially, the sample is magnetized to saturation along the direction (-z) – see in Figure 1(a). For this it is necessary that the applied field $H_{dsat} < 0$ was in absolute value greater than arising in the sample the demagnetizing field, i.e. $|H_{zsat}| > H_d$. Figure 1 (b) shows the ferromagnetic magnetized to saturation along the axis (z) under the action of the field $H_z > |H_d|$. Thus, Figure 1 corresponds to the final states of the ferromagnet. In a given article the process of (a) $\rightarrow$ (b) transition is analyzed.

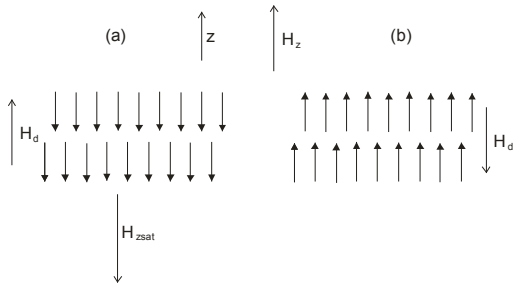

Figure 1: Scheme of the final states of the ferromagnet: (*a*) – up to saturation under a field $H_{zsat}$, and (*b*) – ferromagnet under field $H_z$.

In the third part of the article, the changes in the phase transition of ferromagnet under the action of additional high frequency magnetic field perpendicular to the main field have been considered. In such conditions, SOS of ferromagnet arise. Features of these states have been investigated.

## 2. Phase transition in unidomain ferromagnet

To analyse magnetic phase transition in the ferromagnet with uni-axial anisotropy, we use the Landau–Lifshitz equation [9]:

$$\frac{\partial \mathbf{m}}{\partial t} = \gamma \mathbf{m} \times \frac{\partial W}{\partial \mathbf{m}} + \kappa \left( \mathbf{m} \times \frac{\partial \mathbf{m}}{\partial t} \right) \quad (\kappa > 0) \tag{1}$$

and the following expression for the density of energy:

$$W = \frac{K_1}{2} |m_\perp|^2 - m_z H_z - N_d M_0 m_z. \tag{2}$$

$H_z$ is an external magnetic field directed along the anisotropy axis $Z$ ($H_z > 0$); $\mathbf{m}$ is a non-dimensional vector of ferromagnetism equal (in the absolute value) to $1$; $m_\perp = m_x + i m_y$; $K_1 > 0$, $\gamma = 2\mu_B/\hbar$; $M_0$ is the magnetization of a crystal; initial magnetizaton is along the (-z) direction; $N_d$ is a factor of demagnization for the sample. In present paper $m_z = \pm\sqrt{1 - |m_\perp|^2}$.

Equation (1) can be written as:

$$i\frac{\partial m_\perp}{\partial \tau} = -h_z m_\perp - (1 - 2h_d) m_z m_\perp + \kappa \left( m_\perp \frac{\partial m_z}{\partial \tau} - m_z \frac{\partial m_\perp}{\partial \tau} \right). \tag{3}$$

Here the differentiation is carried out with respect to the dimensionless time $\tau = 2\mu_B K_1 \hbar^{-1} t$; $h = H_z/K_1$; demagnetization field $h_d = N_d \frac{M_0}{K_1}$. From (3), the equations, which define the correspondence between $m_z$ and $\omega$ and the time changes of these parameters are the following:

$$(1 - 2h_d) m_z + h = \omega(1 + \kappa^2), \tag{4}$$

(Since $\kappa^2 \ll 1$, you can neglect this value.)

$$\frac{dm_z}{d\tau} = \kappa \omega (1 - m_z^2). \tag{5}$$

For the energy density relative to initial state we have:

$$e = (1 - 2h_d)\frac{(1 - m_z^2)}{2} - h(1 + m_z), \tag{6}$$

and

$$\frac{de}{d\tau} = -\kappa \omega^2 (1 - m_z^2). \tag{7}$$

In what follows we consider the process of changing of parameters of a ferromagnet in the transition from the initial state when $m_z = -1$. In this process, the energy decreases, respectively $m_z$ increases from the initial value, and precession frequency is changed too. The character of changes in a ferromagnet during the phase transition depends strongly on the shape of the sample. In the following examples, the only cases are used: of a plane and spherical shape. Moreover, the thickness of plane sample is much smaller than the dimensions in other directions, and the anisotropy axis is perpendicular to the plane of a flat sample. In the first case the demagnetization factor $N_d = 4\pi$, while in the second case $N_d = 4\pi/3$.

At first, in Figures 2 to 4 for the plane form of the sample, limit values of main parameters are given as functions of the ratio $M_0/K_1$ for different values of acting field $h$. Initial energy is $e_1 = 0$, final energy is $e_2$. Correspondingly, we have initial $m_{z1} = -1$ and final $m_{z2}$, initial $\omega_1$ and final $\omega_2$. These limiting values are determined from equations (4) and (6).

As can be seen, for a given value of a field, full rotation occurs only at sufficiently small value of $M_0/K_1$. At a higher value of this ratio, the final value $\omega_2 = 0$ and as can be seen in these Figures, the values $m_{z2} < +1$. The field value becomes insufficient to overcome the demagnetizing fields. Figure 3 corresponds to $h = 0$, i.e. when the field $H_{zsatur} < 0$, which magnetizes the sample to saturation, is simply removed. In this case $m_{z2} = 0$. If the current field is negative, i.e. $h < 0$, the value $m_{z2} < 0$, as is shown in Fig.4.

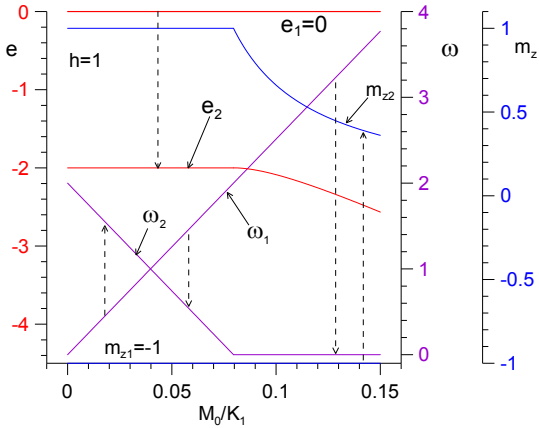

Figure 2: Limit values of main parameters for plane sample vs the relationship $M_0/K_1$ at $h = 1$.

In case of the spherical shape of the sample, dependence of the limit parameters on the value $M_0/K_1$ differs from those shown in Figures 2 - 4 only by the offset of $M_0/K_1$ values, in accordance with decrease of demagnetization factor by three times.

In the range $M_0/K_1 \geq \dfrac{1+h}{2N_d}$, we have the following:

$$\omega_2 = 0, \quad e_2 = \frac{(1-h-2h_d)^2}{2(1-2h_d)}. \tag{8}$$

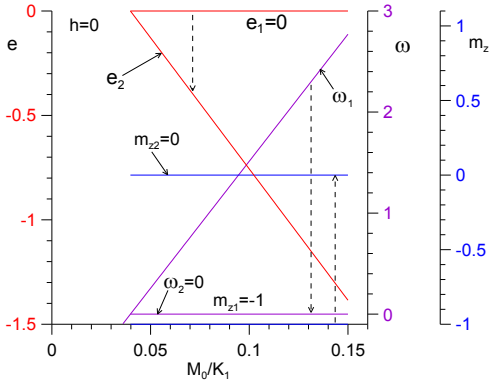

Figure 3: Limit values of main parameters if $h = 0$.

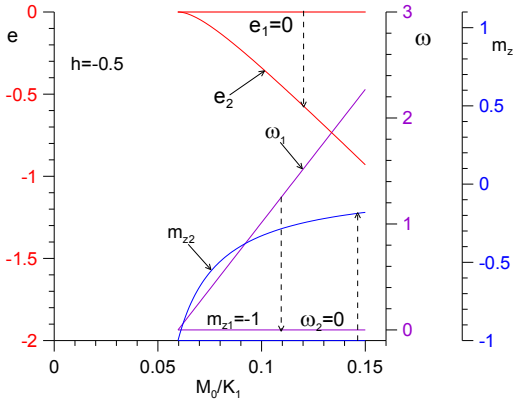

Figure 4: Limit values of main parameters at $h = -0.5$.

Let us consider the time dependence of parameters during the phase transition. In correspondence with equations (4) and (5), the time dependence of $m_z$ can be obtained:

$$\tau = \frac{1}{\kappa} \int_{m_{z0}}^{m_z} \frac{dm_z}{(1 - m_z^2)[h + (1 - 2h_d)m_z]}. \tag{9}$$

In Figures 5 – 8, the time dependences of main parameters for plane sample are presented, according to (9). In all these and in subsequent examples, the dissipation parameter is $\kappa = 5 \times 10^{-4}$. These time changes correctly correspond to dependences of the type shown in Figures 2 – 4. A minimum field, in which a change in orientation occurs, is: $h_{min} = 1 - 2h_d$. For plane sample at $M_0/K_1 = 0.1$, this field equals to $h_{min} = -1.513274$ (an approach to this value can be seen in Figure 8).

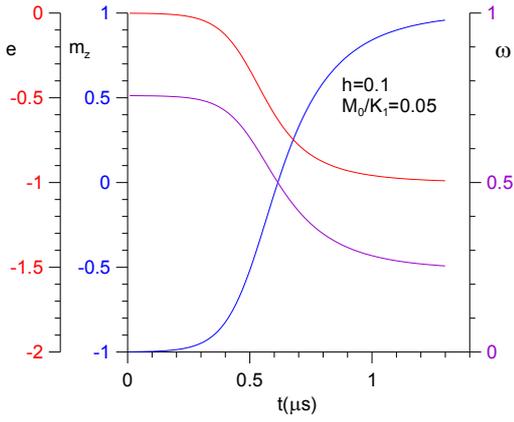

Figure 5: The time dependences of $m_z$, $\omega$ and $e$ for plane sample at $h = 0.1$, $M_0/K_1 = 0.05$.

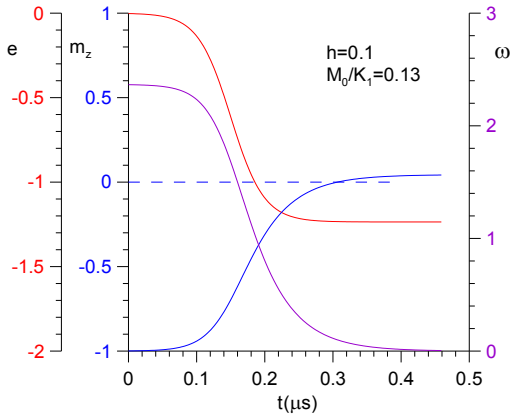

Figure 6: The same as in Figure 5, at $h = 0.1$, $M_0/K_1 = 0.13$.

Figure 7 corresponds to the case when at $h = 0$ the magnetization in final state is perpendicular to the axis of anisotropy. Note that for multidomain sample, zero magnetic field corresponds to the state, when the magnetic moments of domains are directed with equal probability along or against the anisotropy axis.

Thus, for field values within the limits

$$-\left|1 - 2h_d\right| \le h \le \left|2h_d - 1\right| \tag{10}$$

during the phase transition, there is a continuous spectrum of intermediate, seemingly "frozen" states of a ferromagnet. System trends into each of these frozen states (FS) are asymptotical, wherein the precession frequency $\omega \to 0$.

In Figure 9, the time dependences of $m_z$, $\omega$ and $e$ for spherical sample at $h = 0.5$, $M_0/K_1 = 0.065$ are shown.

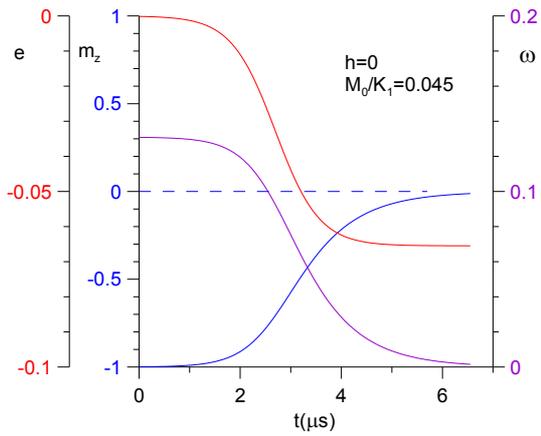

Figure 7: The time dependences of $m_z$, $\omega$ and $e$ for plane sample at $h = 0$, $M_0/K_1 = 0.045$.

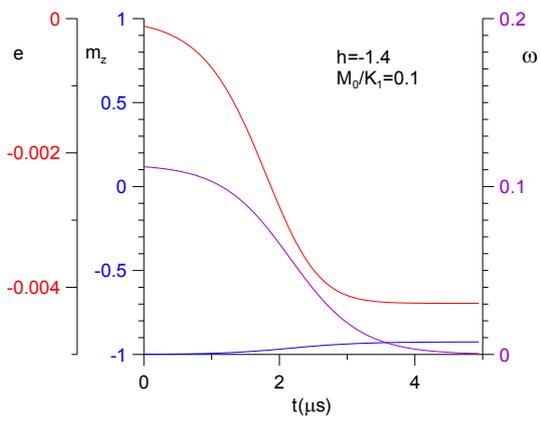

Figure 8: The same as in Figure 7, at $h = -1.4$, $M_0/K_1 = 0.1$

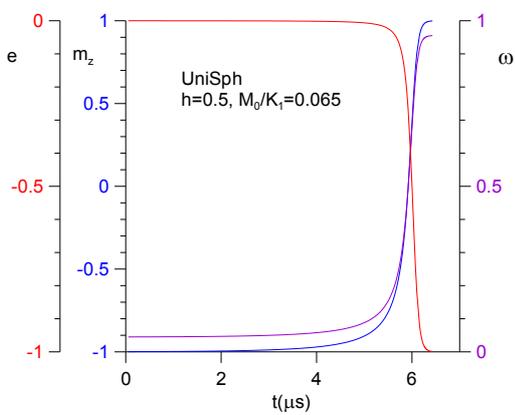

Figure 9: The time dependences of $m_z$, $\omega$ and $e$ for spherical sample at $h = 0.5$, $M_0/K_1 = 0.065$.

In Figure 10, field dependences of energy and $m_z$ value for FS of plane sample are shown, at $M_0/K_1 = 0.1$. For all FS, $\omega \to 0$.

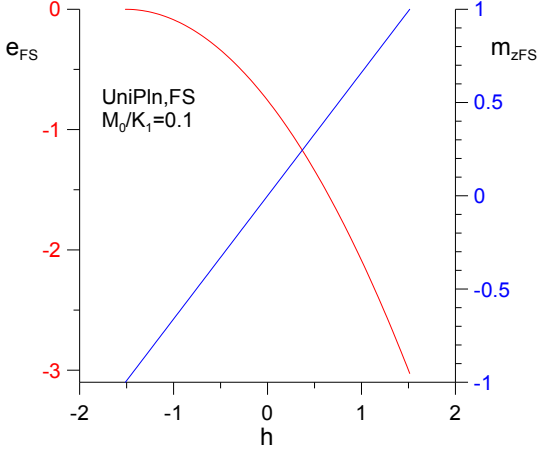

Figure 10: Field dependences of energy and $m_z$ parameter of FS for plane sample at $M_0/K_1 = 0.1$. Here, the frequency for all states is $\omega \to 0$.

## 3. Self-organizing states

Using an additional external magnetic field, we can fix the precession frequency and thereby stabilize the intermediate states of the ferromagnet. If the added periodic field is perpendicular to a main field, and

$$\mathbf{H}_\perp = K_1 h_\perp e^{i\omega\tau}, \tag{11}$$

we can express the magnetic component of magnetization in the form

$$m_\perp(\tau) = p(\tau) e^{i(\omega\tau - \beta(\tau))}, \tag{12}$$

i.e. the precession phase of magnetic moments differs from the phase of periodic field. In this case, the equations for $m_z = \pm\sqrt{1-p^2}$ take the following form:

$$(1-m_z^2)\left[(1-2h_d)m_z + h - \left(\omega - \frac{d\beta}{d\tau}\right)\right] = +\kappa \frac{dm_z}{d\tau} + h_\perp m_z \sqrt{1-m_z^2} \cos\beta, \tag{13}$$

$$\frac{dm_z}{d\tau} = \sqrt{1-m_z^2}\left[\kappa\sqrt{1-m_z^2}\left(\omega - \frac{d\beta}{d\tau}\right) - h_\perp \sin\beta\right] \tag{14}$$

From (6), we obtain expressions for energy density relative to the initial state, together with the energy of interaction with the external field (see, for example, [8]):

$$e = (1 - 2h_d)\frac{(1 - m_z^2)}{2} - h(1 + m_z) - h_\perp \sqrt{1 - m_z^2} \cos\beta \qquad (15)$$

and for the change of this energy connected with dissipation and the action of external periodic field:

$$\frac{de}{d\tau} = -\kappa\left[\frac{1}{1 - m_z^2}\left(\frac{dm_z}{d\tau}\right)^2 + (1 - m_z^2)\left(\omega - \frac{d\beta}{d\tau}\right)^2\right] + h_\perp \sqrt{1 - m_z^2}\,\omega \sin\beta. \qquad (16)$$

The equations (11) – (16) constitute a complete description of the system, including its time transformation. However, in the present paper we consider only dynamic equilibrium state of ferromagnet, i.e. when the decrease of energy caused by dissipation is compensated by energy flow from the external periodic field, i.e. $de(\tau)/d\tau = 0$. Furthermore, in this case $dm_z/d\tau = 0$ and $d\beta/d\tau = 0$. Therefore, for this equilibrium state of ferromagnet, i.e. for self-organizing state (SOS), we obtain the following expressions:

$$\frac{dm_z}{d\tau} = \sqrt{1 - m_z^2}\left(\kappa\omega\sqrt{1 - m_z^2} - h_\perp \sin\beta\right) = 0, \qquad (17)$$

$$\frac{de}{d\tau} = -\omega\sqrt{1 - m_z^2}\left(\kappa\omega\sqrt{1 - m_z^2} - h_\perp \sin\beta\right) = 0. \qquad (18)$$

From these expressions, we obtain the relation:

$$\sin\beta = \left(\kappa\omega\sqrt{1 - m_z^2}/h_\perp\right). \qquad (19)$$

Correspondingly, the corrected equation for SOS takes the following form (instead of (4)):

$$\sqrt{1 - m_z^2}\left[(1 - 2h_d)m_z + h - \omega\right] = m_z\sqrt{h_\perp^2 - \kappa^2\omega^2(1 - m_z^2)}. \qquad (20)$$

From and equations (17) and (18), it can also be seen that the energy compensation and consequently the origin of SOS is possible only if

$$h_\perp \geq h_{\perp\min} = \kappa\omega\sqrt{1 - m_z^2}. \qquad (21)$$

For such a system, the entropy increase connected with dissipation is compensated by the negative flow of the entropy, which is the result of external periodic field. It can be expressed as follows:

$$\frac{ds}{d\tau} = \frac{ds_{diss}}{d\tau} + \frac{ds_{h_\perp}}{d\tau} = 0, \qquad (22)$$

where

$$\frac{ds_{h_\perp}}{d\tau} = -\frac{ds_{diss}}{d\tau} = \frac{1}{T}\frac{de_{diss}}{d\tau} = -\frac{\kappa\omega^2}{T}\left(1-m_z^2\right) < 0. \tag{23}$$

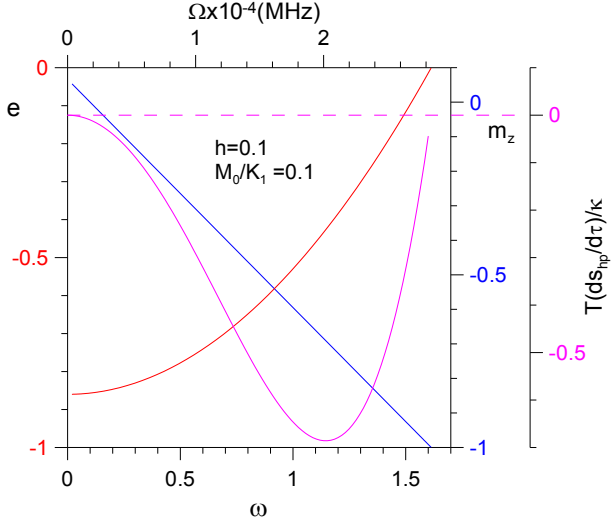

Figure 11: The frequency dependences of energy, $m_z$ value and the negative flow of the entropy due the external periodic field in SOS for plane sample, at $h = 0.1$, $M_0/K_1 = 0.1$. In this case maximum of $m_z$ equals to approx. (+0.05).

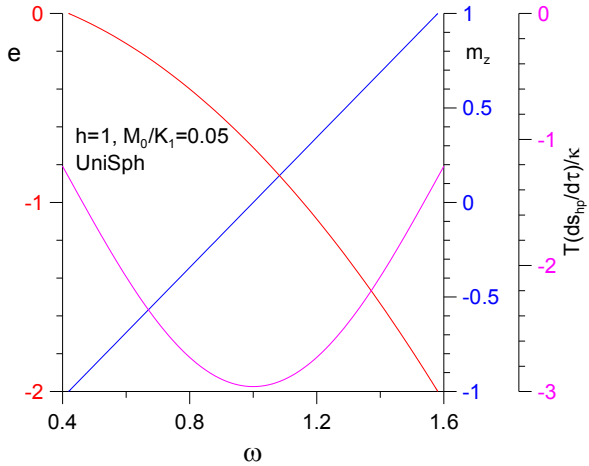

Figure 12: The frequency dependences of energy, $m_z$ value and the negative flow of the entropy due the external periodic field in SOS for spherical sample, at $h = 1$, $M_0/K_1 = 0.05$. In this case, the maximum of $m_z$ equals to (+1).

Two examples of the frequency dependence of energy, value of $m_z$ and the change in entropy for the plane and spherical samples in SOS are presented in Figure 11 and Figure 12. In Figure 13, the frequency dependence of minimum amplitude of periodic field $h_\perp$ for SOS in plane sample, according to (21), is shown.

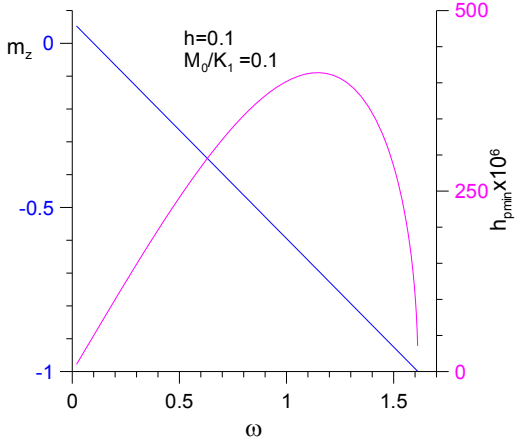

Figure 13: The frequency dependence of $m_z$ value and minimum amplitude of periodic field $h_\perp$ for SOS in plane sample, at $h = 0.1$, $M_0/K_1 = 0.1$.

However, in addition to obtaining such states, we can change these self-organising states. Assume that the precession frequency varies slowly enough. In this case, in Equations (16) - (20) instead of $\omega$, we have $(\omega_0 + \frac{d\omega}{dt}t)$, where $\omega_0$ is the initial frequency. We can in all equations simply replace $(\omega_0 + \frac{d\omega}{dt}t)$ on $\omega(t)$. In result, we obtain the characteristics of self-organising state which depend on time, i.e. $e(t)$, $m_z(t)$ and

$$\sin \beta(t) \cong \kappa \omega(t) h_\perp^{-1} \sqrt{1 - \frac{(\omega(t) - h)^2}{(1 - 2h_d)^2}}, \qquad (24)$$

and, correspondingly, the change of entropy depends on the time too, according to Figure 11 and Figure 12. As a result, changing the frequency of external field, and consequently the energy $e_\perp = h_\perp \sqrt{1 - m_z^2} \cos\beta$ too, we can control the self-organising system, and not only reduce the system energy, but also increase it, reduce the value $m_z$ and return the ferromagnetic in direction to initial phase state.

Further, we can compare the soliton SOS described in [8] with those presented here. Precessing ball solitons of paper [8] may also occur at the first-order transition in a ferromagnet. But their origin is spontaneous and is connected with significant fluctuations in the system configuration. Moreover, the probability of such SOS is strongly dependent on the temperature and the distance from the bifurcation point, in which their energy relative to the initial state is zero.

The SOS presented here, in contrast to [8], are not localized in space, but distributed throughout all volume of the crystal; their appearance is not associated with fluctuations, they do not have a random, probability character, and do not depend on temperature.

## Conclusions

1. A unidomain ferromagnet with uniaxial anisotropy at first-order phase transition under the action of a magnetic field directed along the anisotropy axis has been considered. At certain ratios between the value of acting magnetic field and demagnetization fields, there is a continuous spectrum of "frozen states", which corresponds to different, interconnected values of energy and the deflection angle of the ferromagnetism vector from the anisotropy axis. For these states, precession frequency $\omega \to 0$.

2. At simultaneous action of high-frequency magnetic field perpendicular to the direction of the main field, a self-organizing state of a ferromagnetic arises, in which the ferromagnetic is in dynamic equilibrium. In this equilibrium state, the entropy increase connected with dissipation is compensated by the negative flow of the entropy that is the result of external periodic field.

3. Relations between the main parameters of SOS, i.e. between the values of fields, energy, precession frequency, and the angle between ferromagnetism vector and the anisotropy axis, have been analysed.

4. Changing the frequency of the alternating field, and thereby, the flow of the entropy, can be a continuous method to change all parameters of SOS, including reduction or increase of the system energy.